\documentclass[%
reprint,
superscriptaddress,
 amsmath,amssymb,
 aps,
]{revtex4-1}

\usepackage{graphicx}
\usepackage{dcolumn}
\usepackage{bm}

\usepackage{xcolor}
\usepackage[normalem]{ulem}

\newcommand{\rep}[2]{\textcolor{black}{#2}}

\newcommand{\repp}[2]{\textcolor{black}{#2}}
\newcommand{\rev}[1]{\textcolor{black}{#1}}

\begin{document}

\title{Observation of the spin-orbit gap in bilayer graphene by \\ one-dimensional ballistic transport}


\date{\today}


\author
{L. Banszerus$^{1,2}$, B.~Frohn$^{1}$, T. Fabian$^3$, S.~Somanchi$^{1}$, A.~Epping$^{1,2}$, M. M\"uller$^{1,2}$, D. Neumaier$^{4}$, K.~Watanabe$^5$, 
T. Taniguchi$^5$, F. Libisch$^3$, B.~Beschoten$^{1}$, F. Hassler$^{6}$ and C. Stampfer$^{1,2,*}$
\\
\normalsize{$^1$JARA-FIT and 2nd Institute of Physics, RWTH Aachen University, 52074 Aachen, Germany, EU}\\
\normalsize{$^2$Peter Gr\"unberg Institute (PGI-9), Forschungszentrum J\"ulich, 52425 J\"ulich, Germany, EU}\\
\normalsize{$^3$Institute for Theoretical Physics, TU Wien, 1040 Vienna, Austria, EU}\\
\normalsize{$^4$AMO GmbH, 52074 Aachen,
Germany, EU}\\
\normalsize{$^5$National Institute for Materials Science, 1-1 Namiki, Tsukuba, 305-0044, Japan }\\
\normalsize{$^6$JARA-Institute for Quantum Information, RWTH Aachen University, 52056 Aachen, Germany, EU}\\
\normalsize{$^{*}$Corresponding author; E-mail: stampfer@physik.rwth-aachen.de}
}

\begin{abstract}
\rep{We report on measurements of quantized conductance in gate-defined
quantum point  contacts in bilayer graphene, which show 
ballistic transport with spin polarized conductance of 6~$e^2/h$ at high in-plane magnetic fields.
At the crossings of the Zeeman spin-split subbands of opposite spins, we observe signatures of interaction effects comparable to the 0.7 analog. 
At zero magnetic field the situation seems to be more complex as 
the first subband is already split with a gap that is close to the
expected value for a subband-splitting due to spin-orbit coupling in bilayer graphene, and which can be tuned from 40 to 80~$\mu eV$ by displacement field.  Our results suggest that at zero magnetic field there is an interesting interplay between spin-orbit coupling and electron-electron interaction.}
{We report on measurements of quantized conductance in gate-defined
quantum point contacts in bilayer graphene that allow the observation of subband splittings due to spin-orbit coupling.  The size of this splitting can be tuned from 40 to 80~$\mu eV$ by the displacement field. We assign this gate-tunable subband-splitting to a gap induced by spin-orbit coupling of Kane-Mele type, enhanced by proximity effects due to the substrate.
We show that this spin-orbit coupling gives rise to a complex pattern in low perpendicular magnetic fields, increasing the Zeeman splitting in one valley and suppressing it in the other one. In addition, we observe the existence of a spin polarized channel of 6~$e^2/h$ at high in-plane magnetic field and of signatures of interaction effects at the crossings of spin-split subbands of opposite spins at finite magnetic field.}

\end{abstract}

\maketitle
Bilayer graphene (BLG) represents an interesting platform for mesoscopic transport and
quantum devices. The possibility of tuning the low-energy electronic bands with a perpendicular
electric field is unique to this material~\cite{McCann2013Apr}, \rep{which}{and} allows to open a band gap~\cite{Oos2007Dec,Zhang2009Jun},
to modify band curvatures, and to change the topology of the Fermi surface~\cite{Varlet2014Sep}. As all of this is controlled by external electrostatic gates, it is possible to implement soft-confined one-dimensional channels and quantum dots, where most of \rep{the states in BLG are fully depleted}{the BLG is gapped}. 
Recent technological advancements -- mostly based on the encapsulation of BLG in hexagonal boron nitride (hBN) and on the use of graphite gates -- have enabled the observation of spin and valley states in BLG quantum dots~\cite{Eich2018Jul,Banszerus2018Jun,Eich2018Jula} and of quantized conductance in gate defined quantum point contacts (QPCs)~\cite{ove18j,Overweg2018Dec,Kraft2018Dec,Lee2019Nov}. 
BLG is also interesting for spintronics applications~\cite{Datta90,Gmitra2017Oct,Island2019Jun} because of its weak hyperfine and spin-orbit  interaction~\cite{Yang2011Jul,Ingla-Aynes2016Aug,Leutenantsmeyer2018Sep,Xu2018Sep,Konschuh12}. Spin-orbit (SO) coupling is indeed expected to open a gap of only a few tens of~$\mu$eV in the low energy spectrum of graphene and BLG~\cite{Kane2005Nov,Min2006Oct,Huertas-Hernando2006Oct,Yao2007Jan,Konschuh12}.
\rev{The presence of a SO gap in BLG has been confirmed by the spin lifetime anisotropy in diffusive spin transport~\cite{Leutenantsmeyer2018Sep,Xu2018Sep}. For monolayer graphene, the SO gap has been quantitatively extracted only recently from resonance microwave measurements, giving a value around $40\,\mu$eV~\cite{Sichau2019Feb}.}

\begin{figure}[!thb]
\centering
\includegraphics[draft=false,keepaspectratio=true,clip,width=\linewidth]{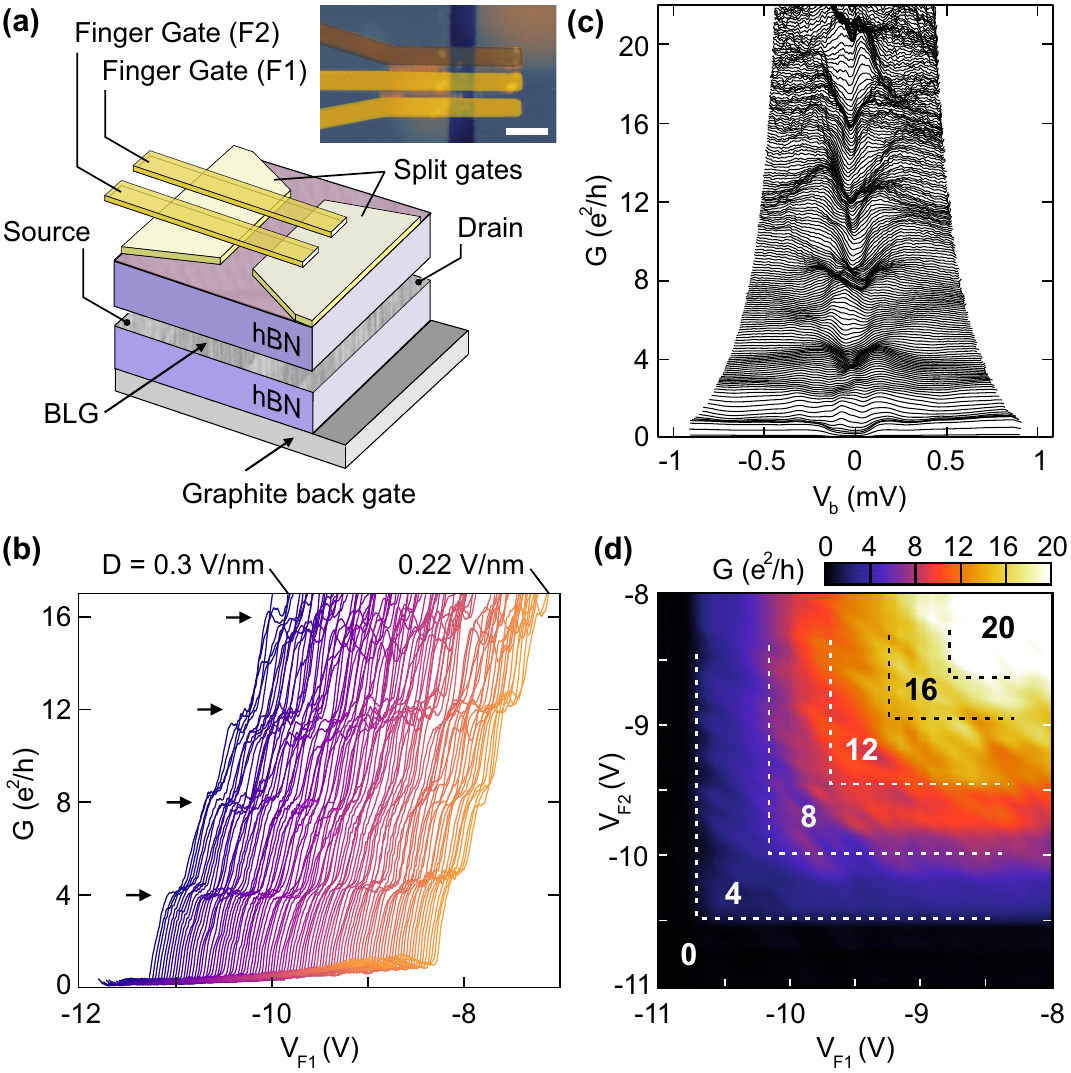}
\caption[Fig01]{ 
\textbf{(a)} Schematic illustration of the device highlighting the hBN/BLG/hBN heterostructure and the various gates. \rep{}{The inset shows an AFM image of the sample. The scale bar is 500~nm (details in Supplementary Material).}
\textbf{(b)} Four-terminal conductance as function of $V_\mathrm{F1}$ for different displacement fields showing steps at multiples of $4\,e^2/h$. \textbf{(c)} Finite bias spectroscopy measurements. Different traces correspond to different values of $V_{\rm F1}$, ranging from -9.4 to -10.6~V.  A clustering of traces at multiples of $4\,e^2/h$ is visible at low bias voltages, vanishing at high bias.  
\textbf{(d)} Conductance through two QPCs in series separated by 260\,nm. The conductance is quantized in multiples of $4\,e^2/h$ and depends solely on the QPC with the lowest number of occupied modes.}
\label{f1}
\end{figure}
\begin{figure*}[!thb]
	\centering
		\includegraphics[draft=false,keepaspectratio=true,clip,width=0.95\linewidth]{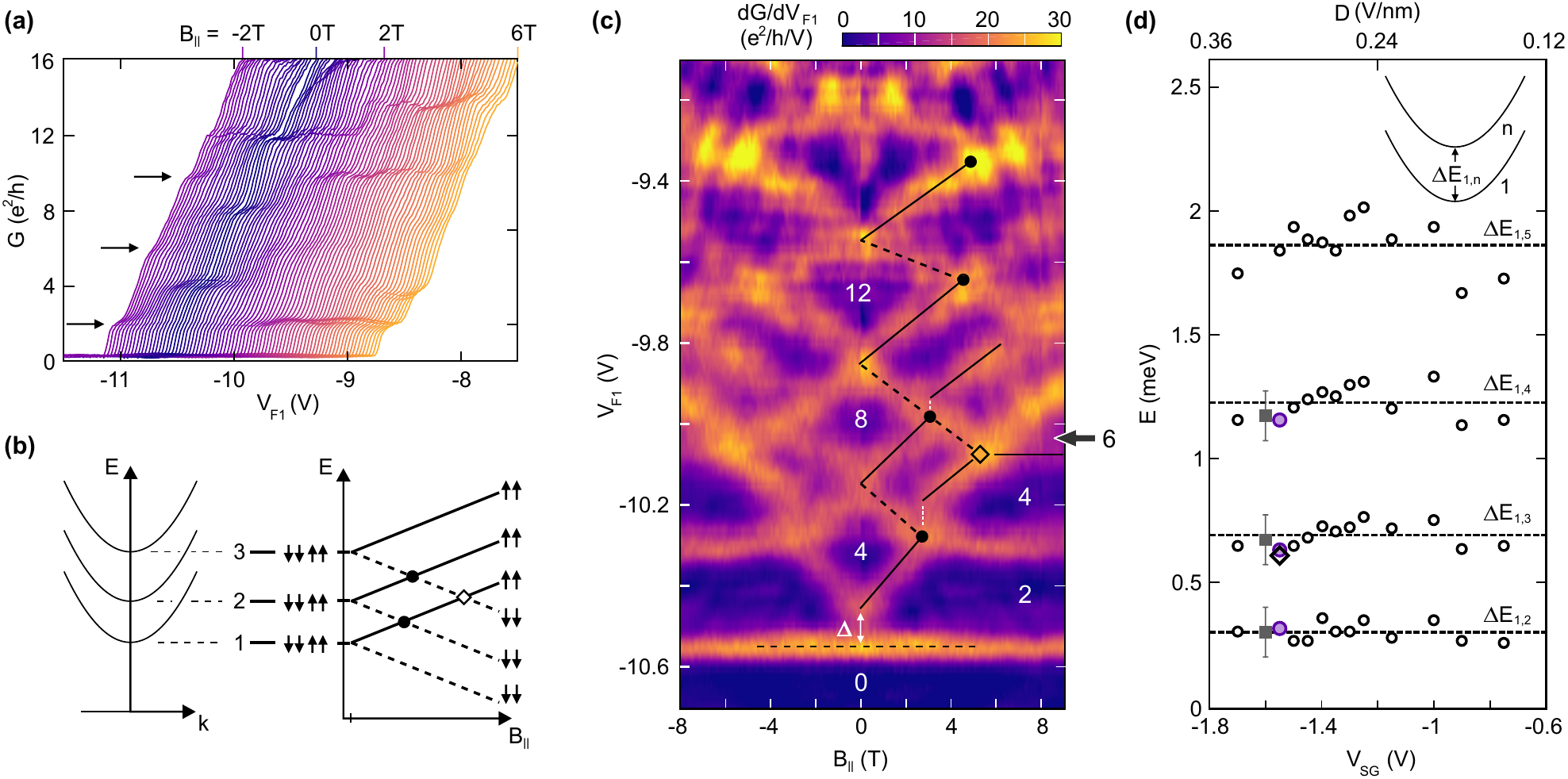}
	\caption[Fig02]{
	\textbf{(a)} Conductance as function of $V_\mathrm{F1}$ for in-plane magnetic fields, $B_\parallel$ between $-2$ and $6\,$T \rev{at $D=0.28\,$V/nm}. Plateaus at $2, 6$ and 10 $e^2/h$ appear. \textbf{(b)} Schematic illustration of the spin splitting of the first three subbands as function of in-plane magnetic field.
	\textbf{(c)} Transconductance d$G$/d$V_\mathrm{F1}$ as function of $V_\mathrm{F1}$ and $B_\parallel$ \rev{at $D=0.28\,$V/nm}. 
	At the points marked by the black dots, the Zeeman energy matches the subband spacing.   
	The rotated square indicates the point where the spin-up states of the first subband cross the spin-down states of the third one, see schematics in panel (b). \rep{}{The numbers indicate the conductance values in units of $e^2/h$.} \textbf{(d)} Energy difference $\Delta E_{1,n}$ between the first and the $n$th subband ($n\le5$),  for different displacement fields. The squared data points originate from finite bias spectroscopy and the round data points from the analysis of the Zeeman splitting, taking $\Delta E_{1,n}=\sum_{i=1}^{n-1}\Delta E_{i,i+1}$. The purple data points are extracted from the data of panel (c). The rotated square represents the energy difference $\Delta E_{1,3}$ as extracted from the position of the rotated square in panel (c). } 
	\label{f2}
\end{figure*}
Probing such small \rep{energies}{energy scales} by transport is challenging, but becomes feasible when having
comparable energy scales in the system, such as small subband spacings in QPCs. 
In addition, small \repp{subband}{} spacings enable observing the crossings of spin-split subbands in parallel magnetic fields. \rep{allowing to observe ballistic spin-polarized transport.
...as well as spin-driven interaction effects such as the 0.7-analog earlier observed in GaAs QPCs~\cite{Graham2003Sep,Berggren2005Mar,Weidinger2018Sep}.}{}
\rep{In this Letter, we report on the observation of  highly spin-polarized currents in a  quantum point contact (QPC) in BLG. By studying a wide QPC with low subband spacings (0.3--0.5\,meV) and comparably large Zeeman energy we are able to observe the crossings of spin-split subbands of opposite spins leading to a regime with six fully spin-down polarized modes at high in-plane magnetic fields. The low subband energies and high energy resolution also allows for the observation (i) of signatures of electron-electron ({\it e-e}\,) interaction at finite magnetic fields (related to the 0.7-analog~\cite{Graham2003Sep,Berggren2005Mar,Weidinger2018Sep}) as well as (ii) of indications of SOC, appearing as a feature at $2\,e^2/h$ due to the splitting of the first subband of the QPC at zero magnetic field.}
{In this Letter we show that by studying a wide QPC with low subband spacings (0.3--0.5\,meV) and comparably large Zeeman energy we can observe the SO gap in BLG, appearing as a feature at $2\,e^2/h$ due to the splitting of the first subband of the QPC at zero magnetic field. 
Studying the Zeeman splitting of the subbands as a function of an in-plane magnetic field, we are able to determine 
the energy scales associated to the system, including the splitting of the first subband at zero magnetic field. The latter ranges from 40 to 80~$\mu$eV, showing 
a monotonic dependence on the displacement field.  
The associated gap can be understood in terms of a substrate-induced enhancement of SO coupling of Kane-Mele type~\cite{Kane2005Nov}. We show that this type of coupling gives rise to a characteristics spin-valley texture in low perpendicular magnetic field, increasing the Zeeman splitting in one of the valley and suppressing it in the other. 
The small energy scales of our device allow also realizing a regime of \repp{unprecedented}{} high spin-polarization, with six fully spin-down polarized modes at high in-plane magnetic fields. Furthermore, at the crossings of spin-split subbands of opposite spins we observe signatures of electron-electron ({\it e-e}) interaction at high magnetic fields.
}

Our device is based on dry-transferred BLG, encapsulated \rep{into}{in} hBN and placed on a graphite back gate, see Fig.~1(a). We use the combination of two Cr/Au split gates (SG) and the graphite back gate (BG) to apply a perpendicular electric displacement field, $D$, that opens up a band gap and depletes large parts of the BLG, defining a quasi-1D channel with a width of around 250\,nm, connecting source and drain contacts. In addition, we place 200\,nm wide finger gates across the channel to locally tune the Fermi energy and thus the number of open modes in the channel~\footnote{The finger gates are isolated from the side gates by a 25\,nm thick Al$_2$O$_3$ layer.}. This forms a QPC below each finger gate. An atomic force microscope image of the device is shown \rep{in Fig.~1 of Ref.~\cite{Banszerus2018Jun}}{as inset} in Fig.~1a.

\begin{figure}[!thb]
\centering
\includegraphics[draft=false,keepaspectratio=true,clip,width=0.94\linewidth]{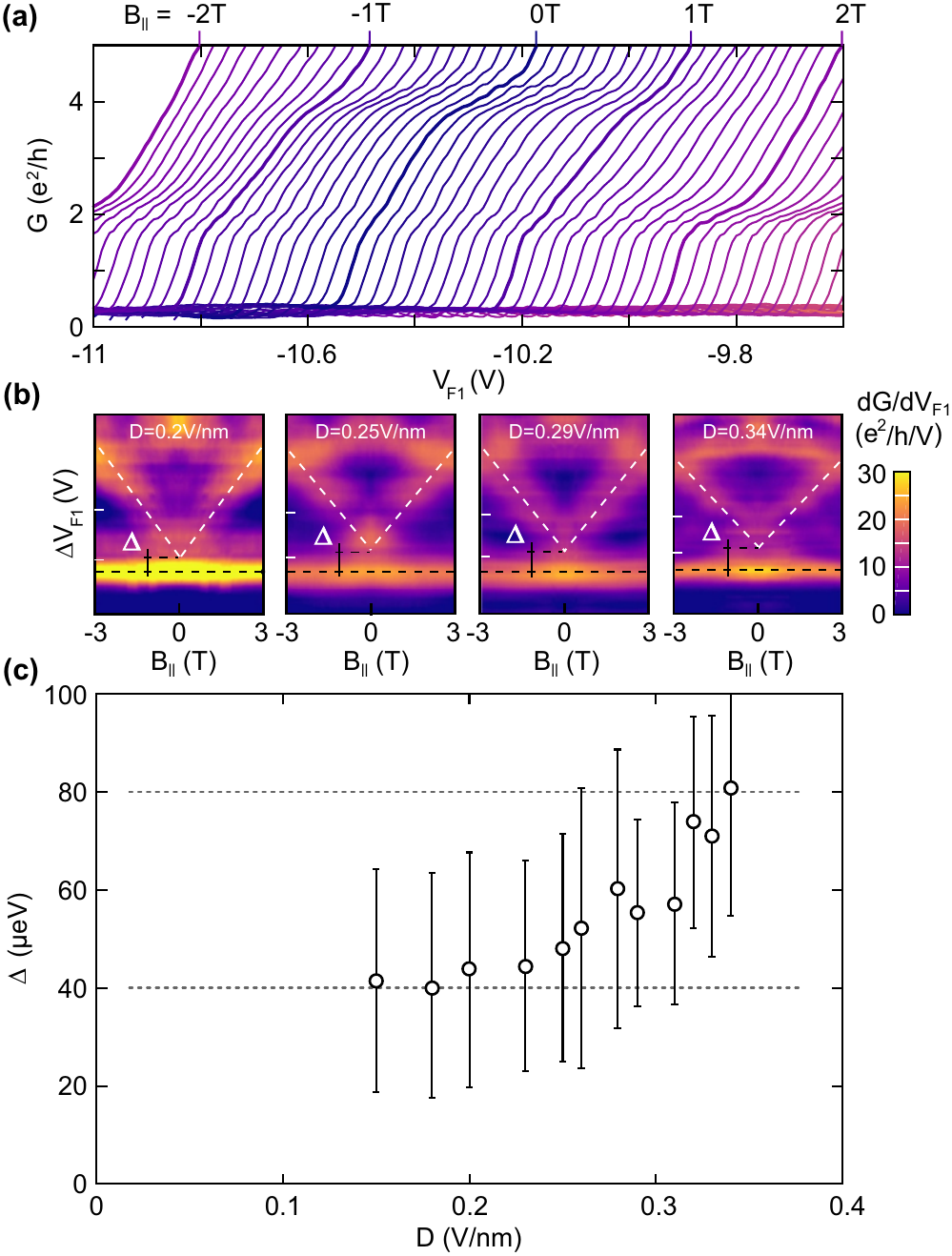}
\caption[Fig03]{
\textbf{(a)} Zoom of Fig.~2(a) around the first conductance step and low $B_\parallel$, showing the presence of a shoulder at $2e^2/h$ even at $B_\parallel=0$. \textbf{(b)} Transconductance as function of $V_\mathrm{F1}$ and $B_\parallel$. The dashed lines mark the evolution of the spin up (white) and spin down bands (black) of the first subband. \textbf{(c)} Extracted energy gap $\Delta$ as function of the displacement field. 
}
\label{f3}
\end{figure}

We perform transport measurements in a He3/He4 dilution refrigerator at a temperature below 30\,mK, using standard lock-in techniques. The four-terminal conductance as function of finger gate voltage $V_\mathrm{F1}$~\footnote{Please note that here the finger gate F1 (F2) corresponds to FG2 (FG3) in Ref. \cite{Banszerus2018Jun}} features \repp{well-developed}{} plateaus at 4, 8, 12 and $16~e^2/h$ for displacement fields ranging from 0.22 to 0.3\,V/nm, see Fig.~1(b). 
The $4\,e^2/h$ step-height indicates four-fold degeneracy (two-fold spin and two-fold valley) and near unity transmission through the QPC. Complete 
current pinch-off is observed for large $D$-fields, i.e. large BLG band gaps. \rep{in the BLG}
Reducing the displacement field increases the leakage current below the split gates, which leads in turn to an increase in the minimum conductance. 
Nevertheless, the height of the conductance steps remains nearly unaffected at $4\,e^2/h$. The near unity transmission through the QPC can be demonstrated more explicitly by using
the second finger gate (F2), i.e. \rep{}{forming} a second QPC placed 60\,nm next to the first one, see Fig.~1(a). 
The conductance of the device as function of both gate voltages $V_\mathrm{F1}$ and $V_\mathrm{F2}$ shows \repp{well-developed}{} steps of multiples of $4\,e^2/h$ and depends only on the QPC with the lowest number of open modes, Fig.~1(d). This observation proves that the two QPCs have unity transmission and that the charge carriers travel ballistically through both QPCs, not thermalizing between them.

To estimate the subband spacings of the quasi-1D system we perform finite bias spectroscopy measurements. Fig.~1(c) depicts the four-terminal differential conductance through the QPC in units of $e^2/h$ as function of the DC bias voltage, $V_\mathrm{b}$, applied between the source and the drain contact (see Fig.~1(a)) for different $V_\mathrm{F1}$ at a fixed $D$-field of 0.32~V/nm. 
The conductance traces bunch at multiples of $4~e^2/h$ for low bias voltages. At higher bias the plateaus smear out revealing energy spacings around 0.3\,meV, 0.4\,meV and 0.5\,meV for the first three subbands (see Supplementary Material and below). These energies are 
a factor 10 smaller than what 
reported in Ref.~\cite{Kraft2018Dec},
because of the large size of our device. Using a hard-wall confinement model for the lowest subband spacing, $\Delta E_{1,2}= 3\hbar^2 \pi^2 / (2m^*W^2)$, where $m^*=0.033 \,m_\mathrm{e}$ is the effective carrier mass in BLG ($m_\mathrm{e}$ is the electron mass),  we estimate the width of the QPC to be $W\approx330\,$nm. This value is in reasonable agreement with the lithographic channel width of 250\,nm.

We investigate the spin structure of the subbands by studying the evolution of the conductance steps as function of an in-plane magnetic field, $B_\parallel$. 
In Fig.~2(a), the conductance is shown as a function of $V_\mathrm{F1}$ for fixed $B_\parallel$-fields ranging from $-2$ to $6\,$T. 
Plateaus at 2, 6,  and 10$\,e^2/h$ emerge with increasing magnetic field (see black arrows in Fig.~2(a)), indicating the lifting of the spin degeneracy of the subbands (see Fig.~2(b)). In Fig.~2(c), we plot the transconductance, d$G$/d$V_\mathrm{F1}$, as function of both $V_\mathrm{F1}$ and $B_\mathrm{\parallel}$. 
The data reveal splittings of all subbands, as seen by the negative and positive slopes of spin-up and spin-down bands. 
Because of the small energy scales of our device, the Zeeman energy matches the subband spacing already for $B_\mathrm{\parallel}$-fields between 2 and 4~T, resulting in the crossing of the spin-up bands with the spin-down bands of the next higher subband (crossings are marked by black dots in Figs.~2(c,b)). 
The feature independent of $B_\parallel$ in Fig.~2(c) corresponds to the spin-down states of the first subband
, which are locked to a finger-gate voltage slightly above $V_\mathrm{F1} = -10.6\,$V because of quantum capacitance effects. 
For $|B_\parallel|>5.8\,$T the spin-up states of the first subband cross the spin-down states of the third subband, giving rise to a regime with six fully spin-down polarized modes ($G \sim 6\,e^2/h$; see black arrow in Fig.~2(c)) -- \repp{an unprecedented}{a very} high-polarization that makes such QPCs interesting for spin polarizers and  detectors in ballistic spin transport devices~\cite{Vila2019Oct}. 

From the data of Fig.~2(c) we \rep{can} also determine the energy spacing between two neighboring  subbands,  $\Delta E_{n,n+1}$. 
At the intersections of the spin-up and spin-down states of adjacent subbands,  
the Zeeman energy $\Delta E_\mathrm{Z}= g \mu_\mathrm{B}B_\parallel$ is equal to the spacing of the two subbands. Using the fact that in graphene and BLG the Lande factor is  $g \sim 2$~\cite{Tans1997Apr,Lyon2017Aug} (as confirmed by direct measurements on our device, see Supplementary Material), we determine the subband spacing $\Delta E_{n,n+1}$  at $B_\parallel=0$. The values determined in this way agree well with those extracted from finite bias measurements
(compare gray squares and purple circles in Fig.~2(d)). The energy difference $\Delta E_{1,3}$ extracted from the position of the rotated square in \rep{panel (c)}{Fig.~2(c)} coincides with the sum $\Delta E_{1,2}+\Delta E_{2,3}$,  further confirming the
consistency of the method.

We \rep{can} investigate the dependence of the subband spacing on the $D$-field by performing measurements such as those shown in Fig.~2(c) but for different $V_\mathrm{SG}$-$V_\mathrm{BG}$ configurations (see Supplementary Material). 
The \rep{}{subband} spacings \rep{between the subbands} appear to be independent of the \rep{applied}{} $D$-field within the margin of the scattering of our data, see Fig.~2(d). 
This indicates (i) that the electronic width of the transport channel is not affected by the different stray-field contributions at different $V_\mathrm{SG}$ values, 
and (ii) that the BLG subband structure 
does not change appreciably when the band gap increases from $\approx 15\,$ to 35\,meV~\footnote{The band gap depends on the displacement field approximately as $D \times$ 80\,meV/(V/nm)~\cite{Zhang2009Jun}}. 

On energy scales below the subband spacing we find a pronounced splitting of the first 
subband at $B = 0$ (marked by $\Delta$ in Fig.~2(c)) leading to a plateau at $2\,e^2/h$, highlighted in Fig.~3(a) (a close-up of Fig.~2(a) around the first conductance step at low $B_\parallel$). The $2\,e^2/h$ feature, which is nearly unaffected by the in-plane magnetic field, can be well understood in terms of a splitting of the first subband caused by SO coupling of Kane-Mele type (see discussion below)~\cite{Kane2005Nov,Min2006Oct,Huertas-Hernando2006Oct,Yao2007Jan,Konschuh12}. 
Notably, we do not observe signatures of {\it e-e} interaction at low magnetic field, which should appear as a conductance step at $\approx 4 \times 0.7\,e^2/h$ at $B_\parallel = 0$, whose height decreases with increasing magnetic field~\cite{Tombros2011Jun}. 
The missing "$0.7$-anomaly" at low magnetic field can be explained by a suppression of the {\it e-e} interaction 
caused by the efficient screening of the 
graphite back gate~\cite{Shapir2019May} (which is only around $25\,$nm away from the 250\,nm wide transport channel), so that at low \repp{magnetic field}{$B$-field} the SO coupling dominates over the interaction. 
We find an overall different situation at 
 high magnetic fields ($B_\parallel > 2$~T), where the SO coupling is quenched due to the magnetic field. 
 We observe a  discontinuous behaviour of the spin-up states at the Zeeman crossings (see jumps highlighted by white dashed lines in Fig.~2(c)), which indicates a spin-splitting caused by {\it e-e} interactions, known as the ``0.7-analog"~\cite{Graham2003Sep,Berggren2005Mar,Weidinger2018Sep}. This effect has been previously observed only in GaAs-based QPCs~\cite{Graham2003Sep}.

\begin{figure}[t]
\centering
\includegraphics[draft=false,keepaspectratio=true,clip,width=\linewidth]{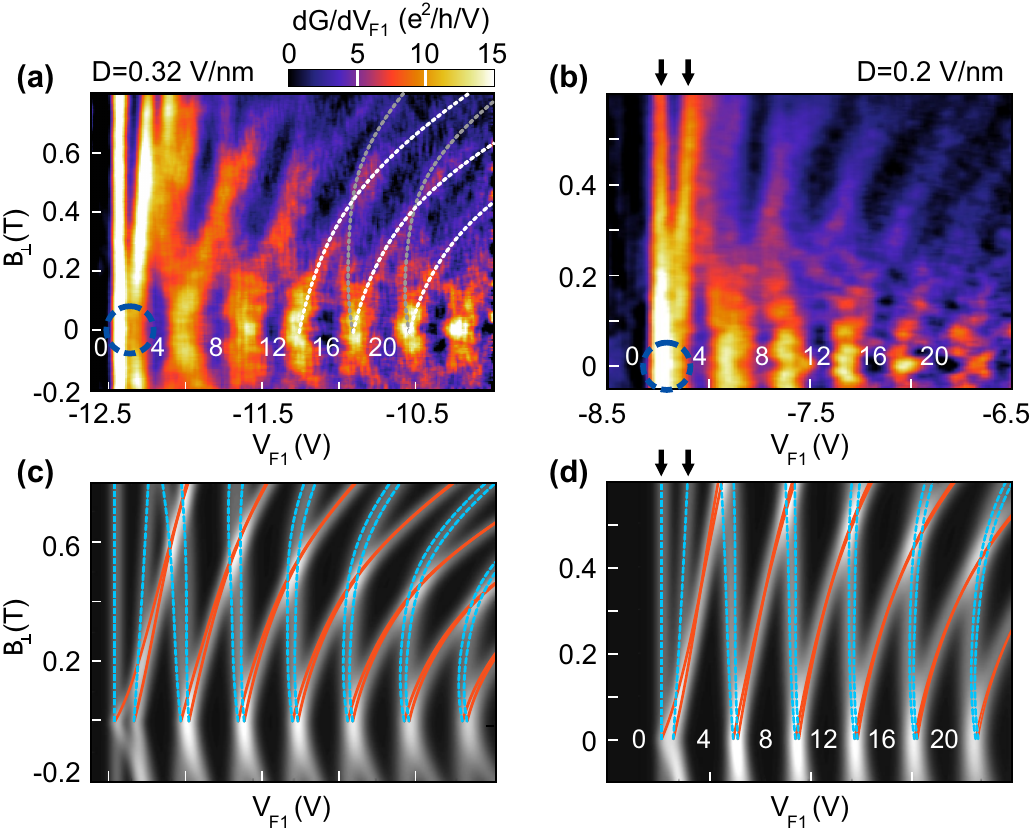}
\caption[Fig04]{\textbf{(a,b)} Transconductance dG/dV$_\mathrm{F1}$ as function of V$_\mathrm{F1}$ and $B_\perp$ for two different values of the displacement field. \textbf{(c,d)} Single-particle model calculations of  dG/dV$_\mathrm{F1}$ \rep{based on the phenomenologically SOC parameters  $\lambda_{\mathrm{lo}}=40\,\mu$eV and $\lambda_{\mathrm{up}}= 80\,\mu$eV (panel c) and $\lambda_{\mathrm{lo}}=\lambda_{\mathrm{up}}=40\,\mu$eV (panel d).}{} The K-valley states are highlighted by dashed blue lines, and the K'-valley states by solid \rev{red} lines.
The gray-scale background shows an approximation for the differential conductance obtained by gaussian smearing. 
}
\label{f4}
\end{figure}
\rep{The presence of the  $2\,e^2/h$-feature indicates the energy splitting of the first subband at $B_\parallel=0$.}{} 
The splitting of the first subband appears clearly in transconductance data such as those of Fig.~2(c) and Fig.~3(b).
The analysis of the subband spacing performed before allows us to \rep{associate an estimated}{estimate the} energy scale \rep{to}{of} the splitting of the lowest subband, $\Delta$, extracted from the transconductance data (see Supplementary Material). 
Performing this analysis for different values of the displacement field, we obtain values of $\Delta$ that range from 40 to 80$\,\mu$eV, with a \repp{nearly linear}{monotonic} dependence on the $D$-field in the observed parameter range, see Fig.~3(c).
This energy scale agrees well with what is expected for the SO gap in graphene and BLG~\cite{Kane2005Nov,Min2006Oct,Huertas-Hernando2006Oct,Yao2007Jan,Konschuh12}, and with the experimental value determined by Sichau {\em et al.} for graphene on SiO$_2$~\cite{Sichau2019Feb}.
\repp{
The observed dependence of $\Delta$ on the displacement field might originate from the fact that SO coupling in BLG is enhanced by proximity effects~\cite{Zollner2019Mar,PrivCom}.
As the latter is influenced by the overlap between the orbitals of graphene and of hBN it likely depends on the $D$-field.}{
The observed dependence of $\Delta$ on the displacement field might originate from the fact that SO coupling in BLG is enhanced by proximity effects (i.e. by the overlap between the orbitals of graphene and of hBN) 
and from details of the $D$-field dependent BLG band structure near the Fermi energy~\cite{Gmitra2017Oct,Zollner2019Mar,PrivCom}.
}

\rep{We speculate that the missing 0.7-feature at low magnetic field can be explained by a small {\it e-e} interaction ($\simeq 1-10\,\mu$eV) such that the SOC dominates over the interaction. At finite in-plane magnetic fields with $B_\parallel\simeq 2-3\,$T, the Zeeman effect quenches the SOC interaction and the {\it e-e} interaction becomes relevant at the first subband crossing leading to the 0.7-analog.}{} Figure~4(a,b) shows the transconductance as function of out-of-plane  magnetic field, $B_\perp$, and finger-gate voltage $V_\mathrm{F1}$. In good agreement with earlier work~\cite{Overweg2018Dec,Kraft2018Dec}, we observe the lifting of the valley degeneracy 
due to nontrivial valley-dependent orbital magnetic moments~\cite{McCann2006Mar} (see dashed lines in Fig.~4(a)), and a characteristic crossing pattern at  increasing magnetic fields. 
The $D$-field dependent splitting of the first subband at $B_\perp=0$ is also \rep{clearly visible}{visible (see dashed circle in Fig. 4(a,b))}.
To reproduce most of this pattern -- including the splitting of the first subband --  we extended the single-particle model developed 
in  Ref.~\cite{Overweg2018Dec,Knothe2018Oct} to take into account also the Zeeman term and the effects of \repp{SOC}{SO coupling} but neglecting {\it e-e} interaction.
\rep{In intrinsic graphene and BLG the dominant SOC term is of Kane-Mele (KM) type~\cite{Kane2005Nov} and couples spin and valley degrees of freedom to preserve time reversal symmetry. 
As for low energies the sublattice and the layer degree of freedom in BLG become equivalent 

we write the SOC
Hamiltonian as}{We write the SO coupling
Hamiltonian as}

$H_{\mathrm{KM}}= \frac 12 [(\lambda_\mathrm{lo}-\lambda_\mathrm{up})\sigma_0 - (\lambda_\mathrm{lo}+\lambda_\mathrm{up})\sigma_z] \tau_\mathrm{z}s_\mathrm{z},$
where $\sigma, \tau$ and $s$ refer to  the layer, 
valley and spin degree of freedom, respectively. This Hamiltonian is of Kane-Mele (KM) type~\cite{Kane2005Nov}. Here we took into account that for BLG the valley and the layer degree of freedom are equivalent at low energies, and we also assume that the proximity-enhanced SO coupling coefficients $\lambda_\mathrm{up,lo}$ are different in the upper (up) and lower (lo) layer of BLG. 

Using an electrostatic simulation of the device to determine the local potential in the BLG (see Supplemental Material), we find overall good agreement between theory and experiment (compare Figs.~4(a,b) with 4(c,d)), by taking phenomenologically $\lambda_{\mathrm{lo}}=40\,\mu$eV and $\lambda_{\mathrm{up}}= 80\,\mu$eV in Fig.~4(c), and $\lambda_{\mathrm{lo}}=\lambda_{\mathrm{up}}=40\,\mu$eV  in Fig.~4(d). 
\repp{We thereby account for the effects that the $D$-field shifts electrons into the upper layer, changes tight-binding hopping parameters, and increases the proximity induced spin-orbit coupling.}
{We  thereby  account for  the  effects  that  the $D$-field localizes electrons on the upper layer,  changes tight-binding hopping parameters, and effectively increases  the  proximity  induced  SO  coupling.}
The calculations reproduce well the splittings of the first subband at zero and finite $B$-field. 
Furthermore, the calculation \repp{also}{} reveal an interesting texture, where the spin-valley coupling
due to $H_{\rm KM}$ at low $B$-fields enhances the Zeeman splitting in one valley (diverging blue \repp{dashed}{} lines in Fig.~4(c,d)) but it suppresses the splitting in the other valley (converging \rev{red} lines).

\rep{To conclude, we have observed the crossings of spin-split 1D subbands leading (i) to highly spin polarized ballistic currents 
 and (ii) to interaction-driven spontaneous spin splittings at the crossing points. The latter is a hallmark of the 0.7-analog structure opening an interesting route for better understanding {\it e-e} interaction effects in BLG quantum wires with a unique valley degree of freedom. This becomes even more interesting as at zero magnetic field the {\it e-e} interaction seems to be in competition with (proximity tunable) SOC, possibly leading to topologically non-trivial ground states.}
 {In summary, we have used the high energy-resolution provided by the confinement effect in a QPC to determine from transport measurements the energy gap due to SO coupling in BLG. The SO gap appears to be tunable between $40\,\mu$eV and $80\,\mu$eV. 
 Controlling the SO coupling is interesting for spintronic applications such as spin-based field effect transistors and spin-orbit valves \cite{Datta90,Gmitra2017Oct,Island2019Jun}, and it might even allow to drive the system in the quantum spin-Hall phase and other interesting topological phases. 
 Furthermore, our measurements in parallel  magnetic field indicate the existence of a regime with six fully spin-down  polarized modes, which indicates the potential of gate-defined QPCs as efficient spin polarizers and detectors in ballistic BLG devices. 
}

\textbf{Acknowledgements:}
We thank J. Fabian, M. Gmitra, A. Knothe, F. Haupt and B. van Wees for helpful discussions.
This project has received funding from the European Union's Horizon 2020 research and innovation programme under grant agreement No 785219 (Graphene Flagship), the Deutsche Forschungsgemeinschaft (DFG, German Research Foundation) under Germany's Excellence Strategy - Cluster of Excellence Matter and Light for Quantum Computing (ML4Q) EXC 2004/1 - 390534769, 
through DFG (BE 2441/9-1 and STA 1146/11-1), the Austrian WWTF Project No. MA14-002, and by the Helmholtz Nano Facility~\cite{Albrecht2017May}. 
Growth of hexagonal boron nitride crystals was supported by the Elemental Strategy Initiative conducted by the MEXT, Japan and the CREST(JPMJCR15F3), JST.

\bibliography{literature}
\end{document}


\title{\bf Supplemental Material for: \\ Observation of the spin-orbit gap in bilayer graphene by one-dimensional ballistic transport}

\date{\today}

\author
{L. Banszerus$^{1,2}$, B.~Frohn$^{1}$, T. Fabian$^3$, S.~Somanchi$^{1}$, A.~Epping$^{1,2}$, M. M\"uller$^{1,2}$, \\  D. Neumaier$^{4}$, K.~Watanabe$^5$, 
T. Taniguchi$^5$, F. Libisch$^3$, B.~Beschoten$^{1}$,\\  F. Hassler$^{6}$ and C. Stampfer$^{1,2,*}$
\\
\normalsize{$^1$JARA-FIT and 2nd Institute of Physics, RWTH Aachen University, 52074 Aachen, Germany}\\
\normalsize{$^2$Peter Gr\"unberg Institute (PGI-9), Forschungszentrum J\"ulich, 52425 J\"ulich, Germany}\\
\normalsize{$^3$Institute for Theoretical Physics, TU Wien, 1040 Vienna, Austria}\\
\normalsize{$^4$AMO GmbH, 52074 Aachen,
Germany}\\
\normalsize{$^5$National Institute for Materials Science, 1-1 Namiki, Tsukuba, 305-0044, Japan }\\
\normalsize{$^6$JARA-Institute for Quantum Information, RWTH Aachen University, 52056 Aachen, Germany}\\
\normalsize{$^{*}$Corresponding author; E-mail: stampfer@physik.rwth-aachen.de}
}

\maketitle
\newpage 

\section{Experimental details}
\subsection{Sample fabrication and measurement technique}
The graphite/hBN/BLG/hBN van der Waals heterostructure was fabricated using a standard dry transfer technique, where the bottom and top hBN have thicknesses of 30~nm and 20~nm, respectively.  One-dimensional Cr/Au contacts were fabricated using electron beam (e-beam) lithography, SF$_6$-based reactive ion etching (RIE) followed by metal evaporation and standard lift-off. Subsequently, the 5$\,\mu$m wide split gates separated by a 250\,nm channel were made by e-beam lithography, metal evaporation (5~nm Cr and 50~nm Au) and lift-off. Next we have deposited 28\,nm of Al$_2$O$_3$ using atomic layer deposition (ALD). As a last step we patterned the finger gates - made out of 5~nm Cr and 70~nm Au (again by a standard e-beam and lift-off process) - such that they cross the channel as depicted in Fig.~S1 and Fig.~1(a) in the main manuscript.
\newline
\newline
All measurements were performed using standard Lock-in techniques in a He3/He4 dilution refrigerator at a temperature below 30\,mK. The conductance through the device was measured using a four-terminal configuration, where all four contacts are located around 10$\,\mu$m away from the QPC, allowing the charge carriers to thermalize. This (quasi two-terminal) configuration allows to measure the resistance through the QPC, while removing the contribution of most of the parasitic resistances. Additionally, a (displacement field dependent) resistance on the order of 600$\,\Omega$ is subtracted from the data to account for the serial resistance of the regions of the 5$\,\mu$m long channel, which are not tuned by the finger gate.  

\subsection{Subband splittings in parallel magnetic field}
In Figure~S2, we present transconductance measurements (d$G$/d$V_\mathrm{F1}$ in units of $e^2/(hV)$) as function of $B_\parallel$ and $V_\mathrm{F1}$ for different displacement fields, i.e. different split gate voltages (see labels in Fig.~S2). These measurements are similar to what is shown in Figure~2(c) of the main manuscript. These data have been used to extract the different subband spacings in Figure~2(d) taking a $g$-factor of $g \sim 2$ (see section 1.3 below). In Figure~S3, we present the data from Figure~2(d) of the main manuscript, highlighting the different conductance values and the spin texture of the observed conductance plateaus.

\subsection{Extraction of the Lande factor $g \sim 2$}
A spin Lande factor of $g \sim 2$ for carbon-based materials such as graphene and carbon nanotubes is widely accepted and has been theoretically predicted and recently experimentally verified for bilayer graphene nanostructures~\cite{Eich2018Jul,Tans1997Apr,Lyon2017Aug}. Since the device studied in this work offers the opportunity to form a quantum dot by creating a n-p-n junction underneath the finger gate (see Figure~S4(a)) \cite{Banszerus2018Jun,Eich2018Jul}, we are able to directly verify $g \sim 2$ for this particular bilayer graphene device. As we have demonstrated with the very same device in an earlier Nano Letter~\cite{Banszerus2018Jun}, excited states can be probed using finite bias spectroscopy measurements (see Figures~S4(b) and S4(c)). In Fig.~S4(c), we plot the transconductance through the device as function of $V_\mathrm{F1}$ and the applied bias voltage $V_\mathrm{b}$. Parallel to the observed Coulomb diamond edge limited by the ground state (GS), an excited state (ES) is well visible (see arrows in Fig.~S4(c)). Next, we measure the transconductance as function of $V_\mathrm{F1}$ and in-plane magnetic field, $B_\parallel$ at a fixed bias voltage of $V_\mathrm{b}=0.15$~meV (see dashed line in Fig.~S4(c)). As shown in Figure~S4(d), the states move linearly with the applied in-plane magnetic field. In particular, we observe the $B$-field dependent splitting of states (see red dashed lines in Fig.~S4(d)). By extracting this splitting in gate voltages as function of $B_\parallel$ and making use of the known conversion from gate voltage to energy (using the data shown in Fig.~S4(c)) we obtain the data points plotted in Figure~S4(e). This data can be directly 
 compared with a Zeeman splitting, $E_Z$ with $g = 2$ given by $E_Z=\mu_B g B_\parallel$, where $\mu_B$ is the Bohr magneton. As observed in Figure~S4(e) these measurements indeed agree well with a Lande factor of $g = 2$ as expected from predictions and previous experiments. 

\subsection{Finite bias spectroscopy - extraction of the subband spacing}

In order to extract the subband spacing of the QPC, we perform finite bias spectroscopy measurements. In Figure~S5, we plot the transconductance, d$G$/d$V_\mathrm{F1}$ in units of $e^2/h$ as function of the applied bias voltage, $V_\mathrm{b}$, and the applied finger gate voltage, $V_\mathrm{F1}$. At several gate voltages, diamond-shaped regions are observed, where the transconductance is close to zero, corresponding to a conductance plateau (see dashed lines). The subband spacing can directly be read-off from the extensions of these zero transconductance regions on the bias axis. The gray data points indicate the determined subband spacing which have been also used as gray data points in Figure~2(d) of the main text.

\subsection{Estimation of the energy $\Delta$ --- splitting of the first subband}

The 250~nm wide bilayer graphene QPC results in relatively small subband spacings compared to the Zeeman energy. This does not only allows for a large spin polarization at high conductance values as discussed in the main manuscript, but additionally allows to use the Zeeman energy as a energy probe for small energies such as the observed splitting of the first subband, $\Delta$. To extract $\Delta$, first the width of of the conductance plateau (in volts of $V_\mathrm{F1}$) is extracted: To do so, we measure the separation of the conductance step from zero to two $e^2/h$ to the intersection of the step from two to four $e^2/h$ of both field polarities at zero in-plane magnetic field. This gate voltage separation is converted to an energy separation using the spin Zeeman effect: from the slope of the step from two to four $e^2/h$ we convert a separation in gate voltage into an energy separation. 

\subsection{Conductance steps for perpendicular magnetic field}
In Figure~S6, we present the conductance as function of $V_\mathrm{F1}$ at perpendicular magnetic fields ranging from $B_\perp$=0~mT to $B_\perp$=500~mT. The applied displacement field is constant at $D=$~0.24V/nm. At these moderate magnetic fields, we are still in the transition between the size quantization and the Landau quantization regime and at magnetic fields, which are too low to observe full degeneracy lifting of the Landau levels. Nevertheless, the observed conductance steps show a step height of 1~$e^2/h$ at $B_\perp$=500~mT. This observation is in agreement with the observed spin and valley texture of the lowest subband (compare with Figure~4 of the main text).

\newpage

\section{Theoretical model system}

To model transport through the electrostatically defined QPC, we first solve the Poisson equation on a 2D plane perpendicular to the QPC channel using fenics \cite{AlnaesBlechta2015a}, including the split, finger and back gates, with the geometry taken from the experimental device. 
We then include the resulting electrostatic potential on both graphene layers in a $k\cdot p$ Hamiltonian to determine the size quantization energies as a function of magnetic field. 
We include a spin-orbit coupling via $H_{\mathrm{KM}}$ as discussed in the main text. 
Scaling from eigenenergies to a voltage axis is done via the density of states (DOS) of gapped BLG, $V \propto \sqrt{E^2 - m^2/4}$, where $m$ is the ($B$-field dependent) size of the gap.

\subsection{Poisson Solver}
    Transport through the electrostatically defined QPC is controlled by the electrostatic configuration of the gates. In a first step, we therefore solve the Poisson equation on a 2D plane perpendicular to the QPC channel. Given the comparatively low density of states of the two-dimensional layer compared to bulk metallic gates, solving the Poisson equation requires a self-consistency cycle to match the DOS of bilayer graphene to the potential given by the Poisson solver, which we implement in fenics \cite{AlnaesBlechta2015a}.
    We simulate a 2D cut through the experimental geometry (Fig.~1(a) of the main manuscript). As gate voltages we take $1.6$~V on the graphitic back gate, $-1.6$~V on the split gate and $6$~V on the finger gate.
    These potentials are used in the $k \cdot p$ Hamiltonian as $V_\mathrm{up}(x)$ (potential on the upper layer) and $V_\mathrm{lo}(x)$ (potential on the lower layer); see Eq.~S1.

    While any reasonable potential landscape leads to qualitatively similar behaviour, we need the Poisson solver for quantitative agreement, meaning
    that level spacings and level crossings agree with the experimental values. 
    Here, the width of the channel strongly influences the eigenvalue spectrum:
    it influences the level spacing directly -- in a wider channel, the level spacing is smaller and the transition from size quantization to Landau quantization happens at lower $B$ fields.
    Larger separation of the split gates leads to a more ``box'' like separation of consecutive quantization steps, with the subband spacing between subsequent subbands increasing quadratically with subband index.
    When the gates are closer together, the channel represents a harmonic potential with equally spaced subbands or even to such a narrow potential that the subband spacing decreases with $n$.
    Changing the channel width by more than a few percent substantially worsens the agreement with experiment. We obtain the best agreement using the experimental channel width of $250$~nm.

\subsection{Simulation}
    Numerically, we solve the $k \cdot p$ Hamiltonian of bilayer graphene.
    The $k\cdot p$ Hamiltonian is discretized in real space and an electrostatic potential from a 2D cut through the QPC channel is added as on-site energy.
    Due to size quantization in the QPC channel the band structure of the Hamiltonian features subbands (i.e. mini-bands).
    The parametric evolution of the band structure with magnetic field is compared with experiment.
   
    The $k\cdot p$ Hamiltonian for one valley of bilayer graphene including spin is given by \cite{Konschuh2012Mar}
        \begin{equation}
             H_{\text{TB}} = \begin{pmatrix}
             V_\mathrm{up}(x) & \gamma_0 \pi^\dagger & 0 & \gamma_1 \\
             \gamma_0 \pi & V_\mathrm{up}(x) & 0 & 0 \\
             0 & 0 &  V_\mathrm{lo}(x) & \gamma_0 \pi^\dagger \\
             \gamma_1 & 0 & \gamma_0 \pi &  V_\mathrm{lo}(x)
             \end{pmatrix} \otimes
             \begin{pmatrix}
             \ket{\uparrow}\bra{\uparrow} & \\
              & \ket{\downarrow}\bra{\downarrow}
             \end{pmatrix}. 
        \end{equation}
    $V_\mathrm{up}(x)$, and $V_\mathrm{lo}(x)$ are the on-site potentials on the upper and lower layer, as determined by the Poisson solver.     
    We also use a Berry-Mondragon potential $(M \sigma_z)$ with a mass term $M$ much larger than the band gap to avoid edge effects at the boundary of our 500~nm wide simulation box.
    The full Hamiltonian including spin and valley is 
        \begin{equation}
        H^{\mathrm{full}}=
          \begin{pmatrix}
             H_{\text{TB}}(\tau = 1) & 0\\
             0 & H_{\text{TB}}(\tau = -1)
             \end{pmatrix}  + H_{\text{Z}}^{\mathrm{full}} + H_{\text{SO}}^{\mathrm{full}},
        \end{equation}
    where the valley index $\tau$ is implicit in $\pi$ and $\pi^\dagger$ (see below) and $H_{\text{Z}}^{\mathrm{full}} =  \mathds{1}_8\otimes\mu_B g s_z B_z$ is the Zeeman contribution. We use the spin-orbit coupling Hamiltonian 
    \begin{equation}
    H_{\text{SO}}^{\mathrm{full}}=  H_{\text{KM}}^{\mathrm{full}} =  \text {diag} \left( \lambda_\mathrm{up},  -\lambda_\mathrm{up},  \lambda_\mathrm{lo} ,  -\lambda_\mathrm{lo} \right) \otimes s_z \otimes \tau_z.
    \end{equation}
    After downfolding this becomes equivalent to the low-energy Hamiltonian (neglecting to write the tensor products explicitly),
    \begin{equation}
        H_{\text{SO}}= H_{\text{KM}} = \frac 12[(\lambda_\mathrm{lo}-\lambda_\mathrm{up})\sigma_0 - (\lambda_\mathrm{lo}+\lambda_\mathrm{up})\sigma_z]\; s_\mathrm{z}\;\tau_\mathrm{z}.
    \end{equation}
    While we use the full $16\times 16$ Hamiltonian $H^{\mathrm{full}}$ (4 orbitals including spin and valley) in the simulations, we refer to the down-folded one in the main text: at low energies, amplitudes on the two orbitals directly on top of each other (coupled by $\gamma_1$) must be small. Consequently, these degrees of freedom can be neglected reducing the Hamiltonian to 
    an $8\times 8$ matrix. 
    
    For small $D < 10$ meV we find a strong $D$-dependence of the spin-orbit gap due to the change in layer polarisation. By contrast, at experimental $D$-values our model Hamiltonian already yields nearly full layer polarized states  (Fig.~\ref{fs7}), in agreement with experiment.
    We conjecture the dependence of the spin-orbit gap on the displacement field is not due to an induced layer polarization, but mainly from an increase of an effective proximity induced spin-orbit coupling. 
    We model these effects by an explicitly displacement-field-dependent spin-orbit coupling parameter $\lambda_\mathrm{up}$.
    We obtain agreement with experiment when choosing a spin orbit coupling on the upper layer $\lambda_\mathrm{up} = 80 \, \mu$eV in one and  $\lambda_\mathrm{up} = 40 \, \mu$eV in the other simulation, and on the lower layer $\lambda_\mathrm{lo} = 40\, \mu$eV.
    
    In terms of the notation in Ref.~\cite{Konschuh2012Mar}, we take the spin orbit parameters on the upper layer $\lambda_{I2}  = \lambda_{I1} =  \lambda_\mathrm{up}$, and on the other layer $\lambda'_{I2}  = \lambda'_{I1} =  \lambda_\mathrm{lo}$.
    All possible other parameters are set to zero.
    Contributions from Rashba spin orbit coupling are not relevant for the low energy regime in the vicinity of the $K$-point.
    
    We do not solve the eigenproblem for each gate configuration separately, but
    use the fixed potentials of a single gate configuration. 
    In reality, the potential configuration changes as the levels populate. Work along these lines is in progress.
    We have checked explicitly that the relation between applied gate voltage and resulting Fermi energy is linear.
    As long as confinement is achieved within the QPC, the numerical values of the gate voltages do not strongly affect the qualitative behavior. 
    A larger potential difference between back gate and split gate only increases the band gap. 
    
    For a realistic potential configuration $V(x)$ along a cut through the QPC there will be both electron states in the confinement region as well as at the same energy hole states under the split gates.
    These states are not connected to the leads.
    Thus, from the eigenproblem solution we select only those states with large probability amplitude in the channel, and from those we separate electron and hole states by the gap induced by the displacement field.
    
    The parameter values for bilayer graphene are $\gamma_0 = 3.16$~eV  and $\gamma_1 = 0.381$~eV \cite{McCann2013Apr}.
    Given the low-energy excitations we aim to describe, we can safely neglect trigonal warping and electron-hole asymmetry parameters.
    Further, $\pi = \tau p_x + \mathrm i p_y = -v_F \mathrm{i} \left(\tau \partial_x + \left( k_y + x B_z \right) \right)$, $\pi^{\dagger} =  \tau p_x - \mathrm i p_y =  -v_F \mathrm{i} \left(\tau \partial_x - \left( k_y + x B_z \right) \right)$, with Fermi velocity $v_F = \sqrt 3 \gamma_0 a_\text{gr} / 2$, graphene lattice constant $a_\text{gr}=2.46$~\AA, valley index $\tau = \pm 1$, and we set $k_y = 0$.
    The perpendicular magnetic field $B_z$ is included via the minimal substitution $\left( p_x +  \mathrm{i} p_y \right) \rightarrow \left( p_x +  \mathrm{i} p_y +  \mathrm{i} e B_z x  \right)$
    These derivative operators are discretized on a staggered lattice \cite{PhysRevD.16.3031} to avoid Fermion doubling \cite{NIELSEN198120}.
    The two components of the Dirac equation for graphene are evaluated on lattices shifted by half a lattice spacing with respect to each other, avoiding spurious highly oscillatory modes.

\bibliography{literature}

\begin{figure}[H]
\centering
\includegraphics[draft=false,keepaspectratio=true,clip,width=0.4\linewidth]{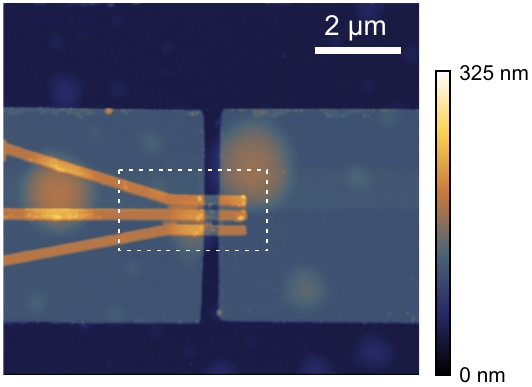}
\caption[FigS6]{Atomic force microscopy (AFM) image of the device. The split gates are shown in light blue and three finger gates (orange) are used to locally tune the potential in the channel between the split gates. The colors correspond to the height information of the AFM scan. The image shown as inset in Fig.~1(a) of the main manuscript is a close-up of this image (see white dashed box), where the contrast of the 3rd finger gate (not used in the experiment) has been reduced.} 
\label{fs6}
\end{figure}

\begin{figure}[H]
\centering
\includegraphics[draft=false,keepaspectratio=true,clip,width=0.95\linewidth]{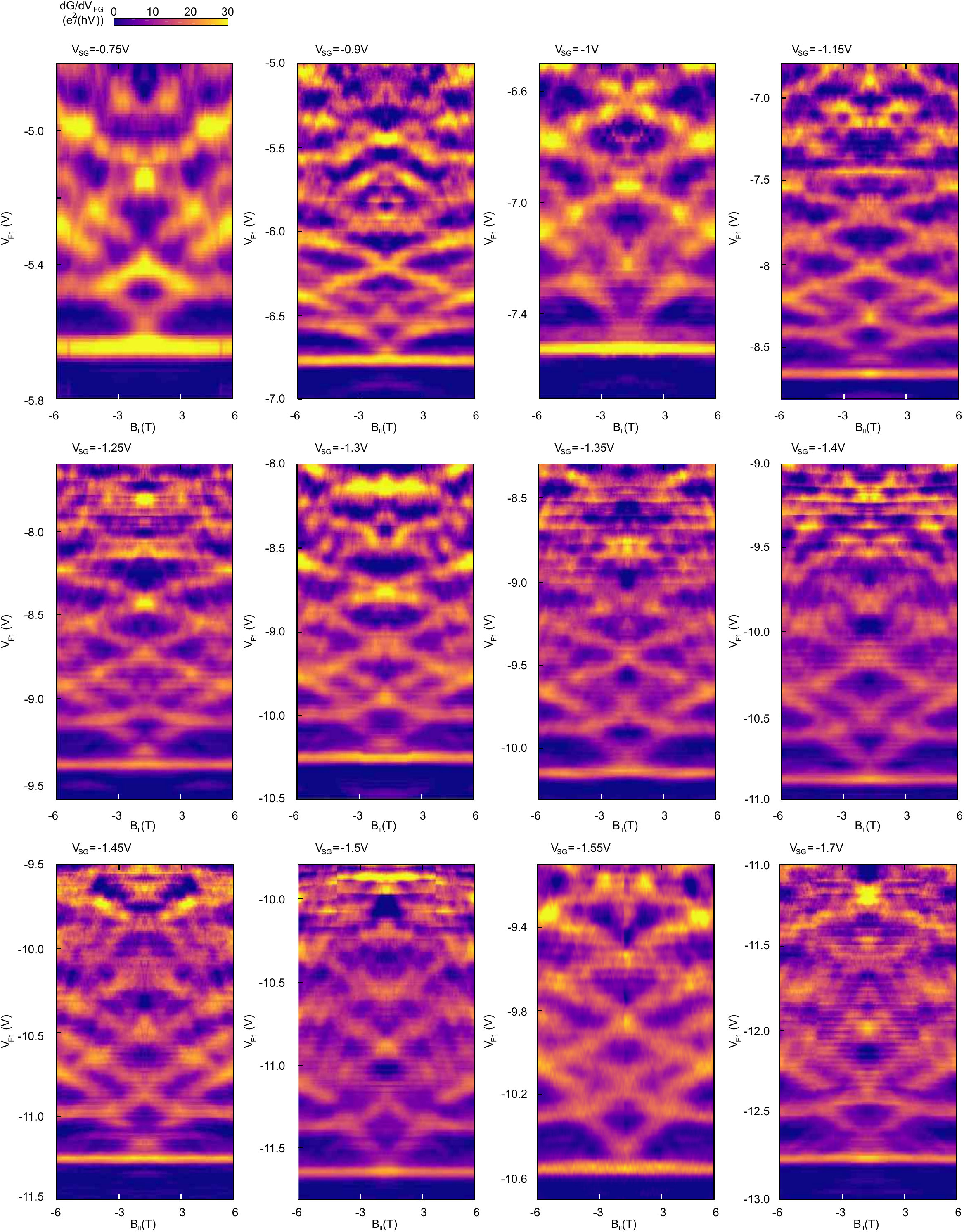}
\caption[Fig01]{Transconductance (d$G$/d$V_\mathrm{F1}$) in units of $e^2/(hV)$ as function of $B_\parallel$ and $V_\mathrm{F1}$ for all displacement fields shown in Fig.~2(d) of the main text.}
\label{fs1}
\end{figure}

\begin{figure}[H]
\centering
\includegraphics[draft=false,keepaspectratio=true,clip,width=0.4\linewidth]{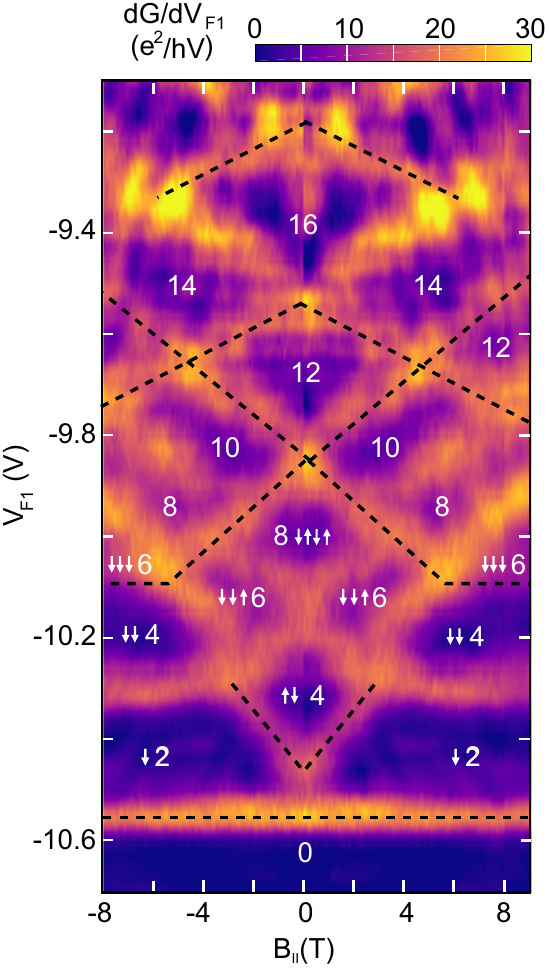}
\caption[FigS2]{\textbf{}Transconductance (d$G$/d$V_\mathrm{F1}$) in units of $e^2/(hV)$ as function of $B_\parallel$ and $V_\mathrm{F1}$ as shown in Fig.~2(c) of the main text. The data is recorded at $V_\mathrm b =0.2$~mV. The conductance in units of $e^2/(hV)$ and the spin texture of the plateaus are given by the labels.}
\label{fs2}
\end{figure}

\begin{figure}[H]
\centering
\includegraphics[draft=false,keepaspectratio=true,clip,width=\linewidth]{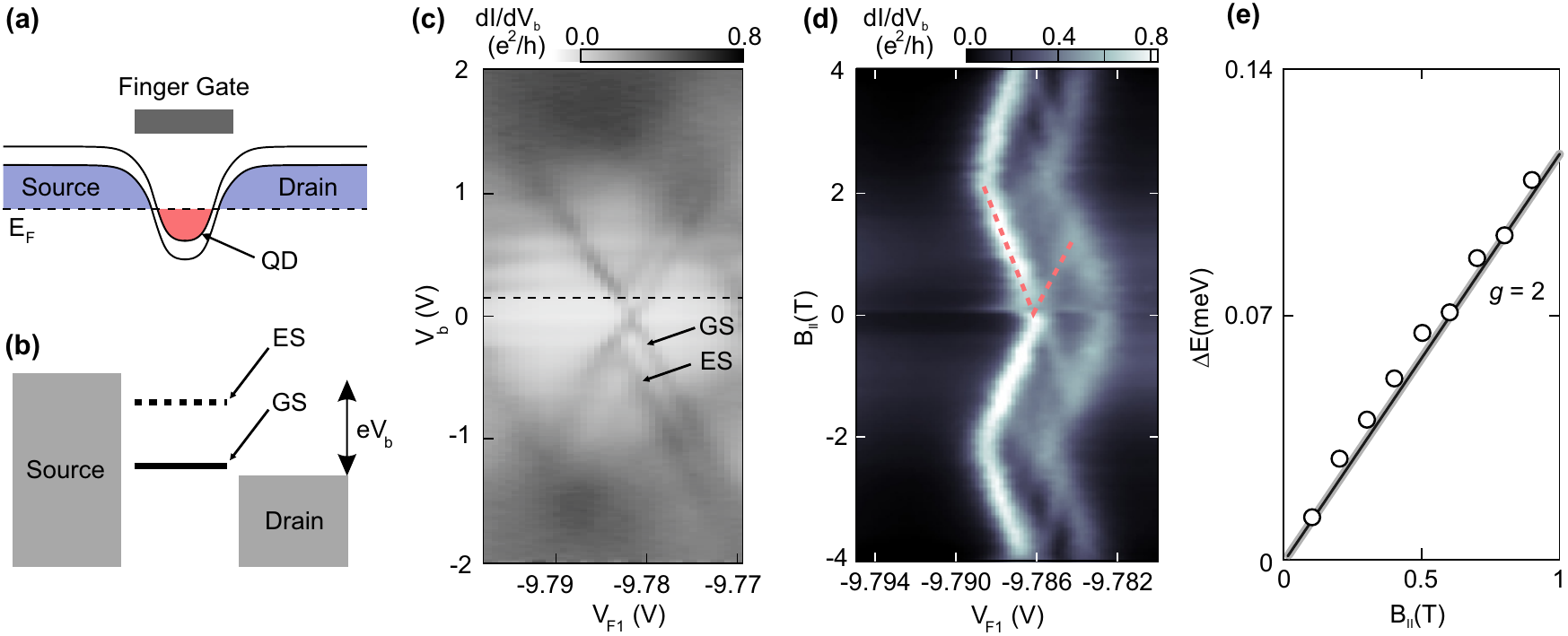}
\caption[FigS3]{\textbf{Excited state spectroscopy on a quantum dot.} \textbf{(a)} Schematic representation of the conduction and valence band along the channel. The quantum dot is formed by a n-p-n junction, induced by finger gate F1.  \textbf{(b)} Schematic illustration of a quantum dot with fixed occupation number. The ground state (GS) and an excited state (ES) reside in the bias window. \textbf{(c)} Finite bias measurements of a quantum dot formed using F1. Apart from the Coulomb diamond formed by the ground state, a parallel excited state is visible. \textbf{(d)} Evolution of the excited state and ground state in a parallel magnetic field at a fixed bias (see dashed line in panel c). \textbf{(e)} Energy splitting between ground state and excited state as function of in-plane $B$-field. The energy splitting increases linearly with $B$ and confirms a spin $g$ factor of 2 for the device. }
\label{fs3}
\end{figure}

\begin{figure}[H]
\centering
\includegraphics[draft=false,keepaspectratio=true,clip,width=0.5\linewidth]{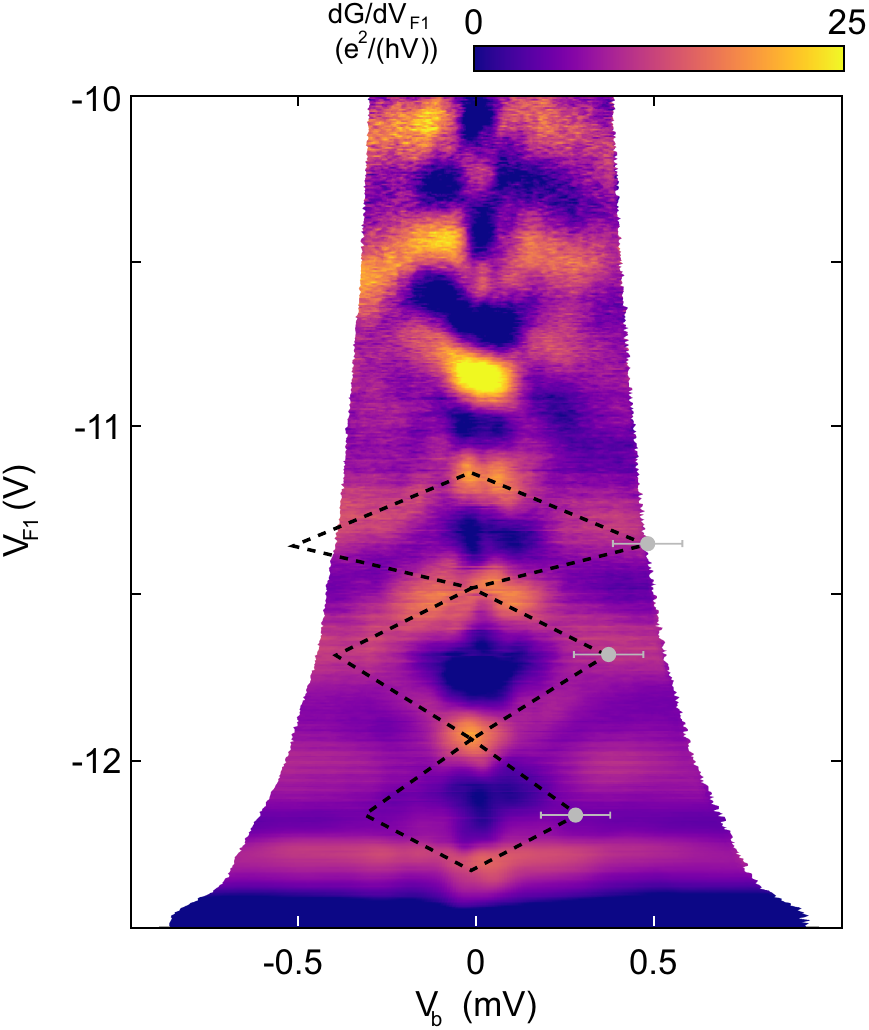}
\caption[FigS3]{Finite bias spectroscopy measurements are used to estimate the subband spacing of the first three modes. The extracted subband spacing is indicated by the black dashed diamonds and the values are used in Fig.~2(d) of the main manuscript. $V_\mathrm{SG}$ is fixed to $-1.6$~V.}
\label{fs3}
\end{figure}

\begin{figure}[H]
\centering
\includegraphics[draft=false,keepaspectratio=true,clip,width=0.4\linewidth]{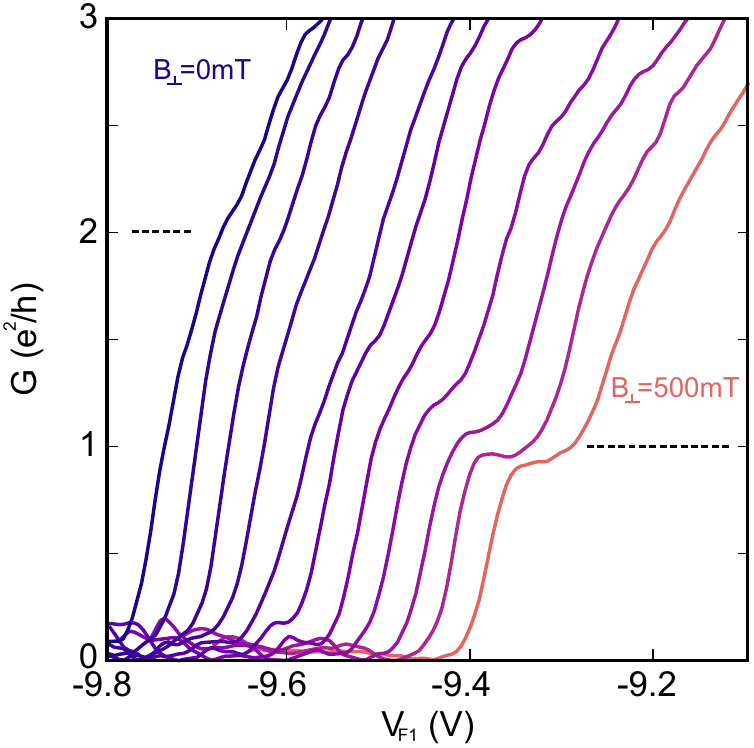}
\caption[FigS3]{Conductance in $e^2/h$ as function of $V_\mathrm{F1}$ at perpendicular magnetic fields ranging from $B=0$~T to $500$~mT. At zero field, the $2\,e^2/h$ plateau is observed. With increasing field a plateau at 1~$e^2/h$ develops, which is in agreement with the model presented in Figs.~4(c) and 4(d) of the main text.}
\label{fs3}
\end{figure}

\begin{figure}[H]
\centering
\includegraphics[draft=false,keepaspectratio=true,clip,width=0.5\linewidth]{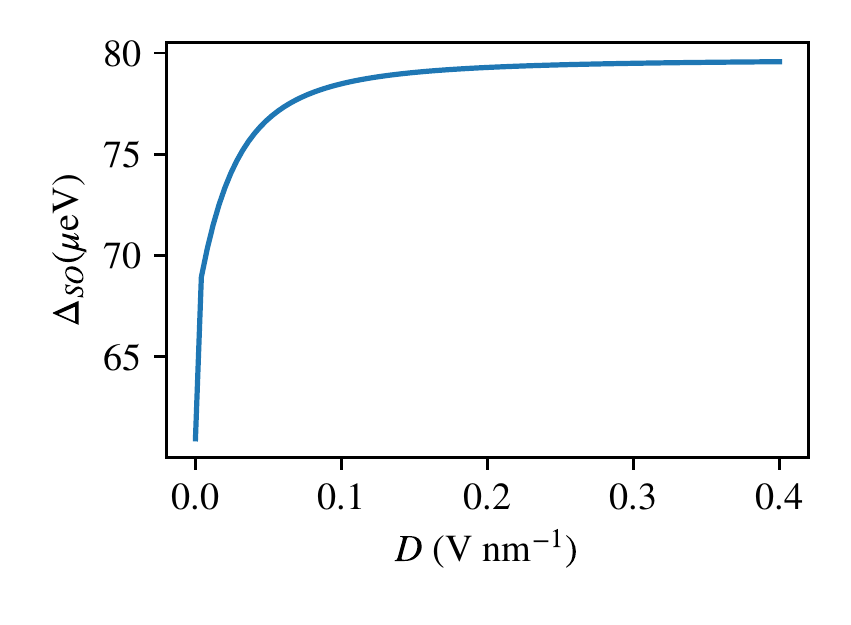}
\caption[FigS7]{
    Evolution of the spin-orbit gap $\Delta_{SO}$ as a function of displacement field $D$ due to layer polarization only (i.e. for $D$-field independent $\lambda_{\mathrm{lo}}$ and $\lambda_{\mathrm{up}}$). The increasing
    layer polarization with increasing $D$ increases the spin-orbit gap for small values of $D$. 
    The spin-orbit coulpling is $\lambda_{\mathrm{lo}}=40\,\mu$eV and $\lambda_{\mathrm{up}}= 80\,\mu$eV on the lower and upper layer respectively.
    At larger $D$, the layers are fully
    polarized and the spin-orbit gap remains approximately constant. 
    } 
\label{fs7}
\end{figure}